\documentclass[aps,prl,twocolumn,showpacs,floats]{revtex4}
\usepackage{graphicx}
\def\etal{{\it et al.}}
\def\prl{{\it Phys. Rev. Lett.}\ }
\def\pr{{\it Phys. Rev.}\ }
\def\w{\omega}
\def\e{\varepsilon}
\def\half{\frac{1}{2}}
\def\squote{}
\def\quote#1#2#3#4{\squote {#1,\ {\sl#2}\ {\bf#3}, #4}.\par}
\def\qquote#1#2#3#4{\squote {#1,\ {\sl#2}\ {\bf#3}, #4};}

\begin{document}
\title{Suppression of Shot Noise in Quantum Point Contacts in the "0.7" Regime }
\author{ A. Golub$^1$, T. Aono$^1$ and Yigal Meir$^{1,2}$}

\affiliation{
$^{1}$ Physics Department, Ben-Gurion University, Beer Sheva 84105, Israel\\
$^{2}$ The Ilse Katz Center for Meso- and Nano-scale Science and
Technology, Ben-Gurion University, Beer Sheva 84105, Israel }
\date{\today}

\begin{abstract}
Experimental investigations of current shot noise in quantum point
contacts show a reduction of the noise near the 0.7 anomaly. It is
demonstrated that such a reduction naturally arises in a model
proposed  recently to explain the characteristics of the 0.7
anomaly in  quantum point contacts in terms of a quasi-bound state, due
to the emergence of two conducting channels. We calculate the shot
noise as a function of temperature, applied voltage and magnetic
field, and demonstrate an excellent agreement with experiments. 
It is predicted that with decreasing temperature, voltage
and magnetic field, the dip in the shot noise is suppressed due to
the Kondo effect.
\end{abstract}
\maketitle

The conductance of  quantum point contacts (QPCs) is quantized in
units of $2e^2/h$ \cite{vanWees,Wharam}. In addition to these
integer conductance steps, an extra conductance plateau around
$0.7 (2e^2/h)$ has been experimentally observed
\cite{Thomas,Kristensen,Reilly,hashimoto,cron}. Recently a
generalized single-impurity Anderson model has been invoked to
describe transport through QPCs \cite{meir}. According to this
model, motivated by density-functional calculations that reveal
the formation of a qusi-bound state at the QPC \cite{hirose}, the tunneling
of a second electron through that state is suppressed
by Coulomb interactions, and is enhanced at low temperatures by
the Kondo effect \cite{kondo}. Thus at temperatures larger than
the Kondo temperature $T_K$, the conductance will be dominated by
transport through the singly occupied level ($G\geq e^2/h$),
growing at lower temperature towards the unitarity limit,
$G=2e^2/h$. Kondo physics has indeed been observed at low
temperature and voltage bias \cite{cron}. The fact that there are
effectively two conductance channels affects not only the
conductance but also the current shot noise. Around conductance of
$G\sim e^2/h$, the model predicts one highly transmitting channel
($T_1\simeq1$) and one poorly transmitting channel ($T_2\simeq0$).
Thus, as the noise is expected to be proportional to the sum of
$T_i(1-T_i)$ over all channels, it should exhibit a dip near that
value of the conductance \cite{moriond}, in contrast with the
traditional view which associate a conductance of $G\sim e^2/h$
with $T_1\simeq T_2\simeq 1/2$ and maximal noise.  A reduction in
the noise through a QPC near $G\sim e^2/h$ has indeed been
observed experimentally \cite{kim,roche,marcus}. The dip was observed
to be quite sensitive to magnetic fields. In
this letter we present a detailed calculation of the noise based
on the above model and demonstrate that it reproduces the
experimental data. The magnetic field dependence arises from two
factors: the dependence of the splitting of the two channels on
the  field, and the quenching of the Kondo effect. Specific predictions on the disappearance of the dip in the
current noise at low temperature, voltage bias and magnetic field,
due to the unitarity limit of the Anderson model are made.

The main theoretical difficulty with calculating the noise is that
the limit of perfect conductance through a given channel is not
accessible via traditional perturbation theory for this
interacting problem. Thus an earlier calculation of the noise through
a Kondo impurity \cite{ournoise} had to rely on more elaborate
methods in order to be extended to lower temperatures. Because of
the additional complexity of the generalized Anderson model,
employed to describe QPCs (see below),  these methods are not
directly applicable. In this work we employ a new type of
perturbation theory, starting from the high magnetic field $B$
limit. In this limit spin-flip processes are suppressed, and the
current and noise can be exactly (and trivially) calculated, 
to all orders in the tunneling, giving rise
to two separate channels. Perturbation in $1/B$ allows us to
follow the contributions and mixing of the two channels. By
comparing to the traditional perturbation theory, around zero $B$,
we are able to interpolate the noise  between the two regimes (see
Eq.~\ref{Gext} below). This formula, which reduces in the known
limits to the obtained perturbative results, allows us to compare
to experiment in the whole magnetic field regime, yielding
excellent agreement with experiment (Fig.~1) and allowing
specific predictions.

 {\it Model Hamiltonian}: The extended Anderson
Hamiltonian, invoked in  \cite{meir} to model the QPC differs from
the usual single-impurity Anderson model in two aspects: (1) the
tunneling amplitude of the first electron into the quasi-bound
state $V^{(1)}$ is larger than that of the second electron
$V^{(2)}$ (see also \cite{guinea}), and (2) both couplings
increase exponentially as the energy of the incoming electron
rises above the QPC barrier, $E_{qpc}$, defined to be
 the zero of energy. This Anderson model can be
transformed into a Kondo Hamiltonian by performing a
Schrieffer-Wolff transformation \cite{SW}

\begin{eqnarray}
H&=& \!\!\sum_{k\sigma \in L,R}
\varepsilon_{k\sigma}c^{\dagger}_{k\sigma}c_{k\sigma}+\!\!\!\!\!\!\!
\sum_{k,k'\sigma \in L,R}
(J^{(1)}_{k\sigma;k'\sigma}-J^{(2)}_{k\sigma;k'\sigma})
c^{\dagger}_{k\sigma}c_{k'\sigma}
\nonumber\\
 & +&2 \!\!\!\sum_ {k,k'\sigma\sigma' \in L,R}
(J^{(1)}_{k\sigma;k'\sigma'}+J^{(2)}_{k\sigma;k'\sigma'})
c^{\dagger}_{k\sigma}\vec{\sigma}_{\sigma\sigma'}
c_{k'\sigma'}\cdot \vec{S} \nonumber
\end{eqnarray}
\begin{equation}\label{int}
    J^{(i)}_{k\sigma;k'\sigma'}=
\frac{(-1)^{(i+1)}}{4}[\frac{V^{(i)}_{k\sigma}
V^{(i)}_{k'\sigma'}}
 {\varepsilon_{k\sigma}-\varepsilon^{(i)}_{\sigma}}+
 \frac{V^{(i)}_{k\sigma}V^{(i)}_{k'\sigma'}}
 {\varepsilon_{k'\sigma'}-\varepsilon^{(i)}_{\sigma'}}]
\end{equation}
where $c^{\dagger}_{k\sigma}(c_{k\sigma})$ creates (destroys) an
electron with momentum $k$ and spin $\sigma$ in lead L or R,
$\varepsilon^{(1)}_{\sigma}=\varepsilon_{\sigma}$ and
$\varepsilon^{(2)}_{\sigma}=\varepsilon_{\sigma}+U $, where
$\varepsilon_{\sigma}$ is the energy of local spin state $\sigma$
and  U is the on-site interaction.  $\vec{S}$ is the local spin
due to the bound state. The potential scattering term (first
line), usually ignored in Kondo problems, is crucial here, as it
gives rise to the large background conductance at high
temperature. The magnetic field B, defining the $z$-direction,
enters the problem via the Zeeman term, $S_z B$. The exponential
increase of the couplings is modelled, for simplicity, by a Fermi
function $f_{FD}(\e)=1/(1+\exp(\e))$, leading to a chemical-potential
dependence
 of the spin-scattering
matrix elements,
\begin{equation}\label{step}
    J_{\e_{k}\e_{k}-B\uparrow\downarrow}^{(i)}=\frac{(-1)^{(i+1)}{V^{(i)}}^2}{4}[\frac{f_{FD}(\frac{-
    \e_{k}}{\delta})}{\e_{k}-
    \e^{(i)}_{\uparrow}}+\frac{f_{FD}(\frac{B-\e_{k}}{\delta})}{\e_{k}-B-
    \e^{(i)}_{\downarrow}}].
\end{equation}
(In the above and in the following $B$ and $T$ denote the corresponding energies,  $g\mu_B B$ and $k_BT$, respectively, where $k_B$ is the Boltzman cosntant, $\mu_B$ is the Bohr magneton,
and with the appropriate $g$-factor.) For 
$B<<\epsilon_F$ we can ignore the magnetic field dependence of
these matrix elements.

 {\it Current noise}: The current noise is defined via the
current-current correlation function \cite{buttiker}
\begin{equation}\label{s}
    \hat{S}(t,t')=\frac{1}{2}(<I(t)I(t')>+<I(t')I(t)>)
\end{equation}
Under stationary conditions  the noise is a function of $t-t'$ and
here we consider only the steady state, zero frequency component of 
 the noise power $S(\w=0)$. The calculation of the
noise, detailed below, consists of the following steps: (a) An
exact solution for very large $B$, where spin-flip processes are
suppressed, for the conductance $G_\infty$ and noise $S_\infty$
(Eq.~\ref{Ginfty}). (b) Expansion to second order in the spin-flip
processes, for arbitrary value of the coupling $J^{(1)}$ and small
value of $J^{(2)}$, yielding $S_B$ (Eq.~\ref{Sb}) (and $G_B$, via the
fluctuation-dissipation theorem). (c) Since the Kondo terms appear
at higher order in perturbation theory, we add the third order
terms in $J^{(2)}$, $G_3$ and $S_3$. (d) We calculate the noise,
at small B, using the traditional expansion in $J^{(i)}$
\cite{appelbaum}. (e) We derive a simple and intuitive
interpolation formulae, for both the conductance and the noise,
that reduce to the obtained expansions in the two limits of small
and large magnetic field. The resulting noise $S$
 and Fano factor $S/I$ are depicted in Fig.~1 and compared to experiments. 

{\it Detailed calculation}: The calculation is carried out using
the non-equilibrium Keldysh Green function approach
\cite{kamenev}.  In this approach there are three independent
Green functions which can be expressed in the terms of the
retarded, advanced and the "Keldysh" Green function, $G^K(\w)$.
For the two leads, the unperturbed Keldysh Green functions are
$g_{k\in L,R,\sigma}^K(\w)=-2\pi i\left(1-2f_{L,R}(\w)\right)$,
where $f_{L,R}(\w)=f_{FD}(\w\pm eV/2)$ are the respective
distribution functions in the leads, which depend on the voltage
difference $V$. It is more convenient to work in the symmetric and
anti-symmetric combinations of the two leads  $g^K
_{\pm}=g^K _{L}\pm g^K _{R}$.

When the magnetic field is large the exchange
part of the Kondo Hamiltonian can be neglected. Therefore $S_z$
can be treated as a conserved "classical" parameter. In this case, averaging over
 $S_z$, one can calculate the conductance and noise exactly,
\begin{eqnarray}
G_\infty&=&\frac{e^2}{h}(T_1 +T_2) \label{Ginfty} \\
S_\infty&=& \frac{e^2}{h}\sum_i[eV \coth(\frac{eV}{2T}) T_i( 1-T_i )+2T
T^{2}_{i}]\nonumber
\end{eqnarray}
Where $T_{1,2}$, the transmission probabilities for the two channels, are expressed
in terms of the coupling constants of Kondo Hamiltonian
$g_i=4\pi\nu J^{(i)}$,
\begin{equation}
T_i=\frac{g_i^2}{1+g_i^2}\label{TH}
\end{equation}
In the large coupling limit the transmission probabilities go to unity.
Since, as function of energy, $g_1$ first increases to a large value, while
$g_2$ becomes large only when $\e_F=\e_0+U$, then, for large magnetic fields, as a function of gate voltage,
 the conductance, in units of $2e^2/h$, will first rise to $\half$ and then to unity.
Concurrently, the shot noise, the first part of $S_\infty$, will have a dip at  the first conductance  plateau, in
agreement with experiments (Fig.~1).

As the magnetic field decreases, the exchange terms in the
Hamiltonian have to be taken into account, influencing and mixing
the contributions of  $J^{(1)}$ and $J^{(2)}$ to the conductance
and the noise. As the magnetic field is still large, we can expand
the conductance and noise to second order in the spin-flip
processes, still allowing infinite order in  $J^{(1)}$ in the non
spin-flip processes. The resulting non-equilibrium noise is a
function of applied voltage and also depends on the
non-equilibrium magnetization \cite {parc} $M(B,T,V)$. The latter
is reduced to its equilibrium value
$M_{eq}=<S_z>=(-1/2)\tanh(B/2T)$ if $B>V$. The resulting
additional contributions to the  noise $S_B$ and the linear response conductance $G_B$ (obtained from the noise via the fluctuation-dissipation theorem, $G=S(V\rightarrow 0 )/(2T) $) in this limit are
\begin{eqnarray}
S_B&=&\frac{e^2}{2h}(g_1
    +g_2)^2[\frac{m_1+m_2}{2}(A_{+} +4BM)\nonumber\\
&\ &\ \ \ \ \ \ - \ m_1 m_2(1+g_1
    g_2)A_{-}]\label{Sb}\nonumber\\
G_B&=& \frac{B}{2T
\sinh\frac{B}{T}}(m_1 +m_2)(g_1+g_2)^2
\end{eqnarray}

where
\begin{eqnarray}
    A_{\pm}&= &B\coth(B/2T) \pm
    \frac{1}{2}\left[ B_{+}\coth(B_{+}/2T)\right.\nonumber\\&\ &\ \ \ \ \ \ \ \ \ \ \ \ \ \ \ \ \ \ \ \left. +B_{-}\coth(B_{-}/2T)\right].
\end{eqnarray}
Here $m_{1,2}=1/(1+g_{1,2}^2)$ and $B_{\pm}=B \pm eV$. In equation
({\ref{Sb}}) $g_2$ was considered small. The nonequilibrium
magnetization is given by \cite{parc} $M=-B/A_{+}$.  In the limit
of small $g_{1,2}$, equation ({\ref{Sb}}) reduces to the zero
frequency current-current correlation function obtained in
\cite{parc}. Note that the corrections to the infinite field
limit, due to spin flips, depend on $\coth(B/2T)$, and thus
decrease exponentially with increasing the ratio $B/T$.

We note that the conductance, to this order, can be written as the expansion
of an expression similar to that of Eq.(\ref{Ginfty}), with $g_i^2$ in Eq.(\ref{TH})
replaced by $\tilde{g_i^2}$, with
\begin{equation}
\tilde{g_i^2}\equiv g_i^2 + \frac{B}{T \sinh\frac{B}{T}}
\frac{(g_1+g_2)^2}{1+(g_1+g_2)^2} \label{Gext}.
\end{equation}

Note that even though $g_2^2$ is small, $\tilde{g_i^2}$ can become
substantial at smaller magnetic field due to higher order processes involving
$g_1^2$. Thus the second channel will also contribute to transport, raising the
 conductance plateau from its value of $0.5\times2e^2/h$ at large magnetic field. This is
 consistent with the observation that the value of "0.7" plateau usually does not
drop experimentally below $0.6\times2e^2/h$.

The second order spin-flip processes do not give rise to Kondo
physics. For this one has to go to third order in the $J$'s. As
argued in \cite{meir}, the Kondo effect will be dominated by the
$J^{(2)}$ term. Due to the step-like increase of the couplings,
the bottom of the band, $E_{qpc}$, is effectively very close to
the Fermi energy, and thus only virtual processes that involve the
empty states will renormalize the couplings. This will be
manifested in the logarithmically divergent terms, arising from
the third order processes. The $J^{(1)}$ terms will involve
integrals over the small region between $E_{qpc}$
 and $\e_F$. The $J^{(2)}$ terms, on the other hand, will involve
integrals from $\e_F$ to the upper band edge $D$ (or to $U$), and
these give rise to the Kondo effect.

These logarithmic contributions only appear  in the Kondo regime
$\e_0 +U>\e_F >\e_0$.
  A lengthy calculation yields the Kondo contribution to the
  noise $S_K$ and conductance $G_K$,
\begin{eqnarray}
    S_K &=&\frac{e^2}{h} \frac{ g_2
    ^3}{\pi}\Big\{\coth(\frac{B_{+}}{2T})\big[F(B)
    +F(B_+)+F(eV)\big] \nonumber \\
     &+&\coth(\frac{B_{-}}{2T})\big[F(B)
    +F(B_{-})-F(eV)\big]
  \nonumber \\
    &+&\coth(\frac{eV}{2T})\big[F(B_{+})-
    F(B_{-})\big]\nonumber \\
   &+&   2 M \big[2 F(B) + F(B_{+}) + F(B_{-} \big] \Big\}\label{SK}
\end{eqnarray}
\begin{equation}
G_K=2\frac{g_2 ^3}{\pi} (1+\frac{2 B}{T
\sinh\frac{B}{T}})\ln\frac{D}{\sqrt{B^2+T^2}} \label{GK}
\end{equation}
 where $F(x)=x \ln(D/\sqrt{x^2+T^2})$.  For $B>eV$ we can apply the expression
  for the equilibrium magnetization $M_{eq}$.

At low temperature (and zero magnetic field) the logarithm contributions to the noise and conducatnce will
diverge, signalling the onset of the Kondo effect below  the Kondo
temperature $T_K\simeq U\exp(-\pi/g_2)$.
 Using the renormalization-group approach, one can sum up
the most divergent logarithms in the higher order Kondo
contributions (Eq.\ref{GK}). Separating the contribution to this Kondo series from the leading terms
and summing up the series leads to the final expression for the total conductance,
\begin{eqnarray}
G_{tot}&=&\frac{e^2}{h}(\hat{T_1} +\hat{T_2})\label{Gtot} \\
\hat{T_i}&=&\frac{\hat{g_i^2}}{1+\hat{g_i^2}},\nonumber
\end{eqnarray}
with
\begin{eqnarray}
\hat{g_1^2}&\equiv& \tilde{g_1^2}\nonumber \\
\hat{g_2^2}&\equiv& \tilde{g_2^2}+g_2^2(\frac{1}{2}-\frac{B}{T
\sinh\frac{B}{T}}) + G_2^{RG},
\label{gi}
\end{eqnarray}
and
\begin{equation}
G_2^{RG}=\frac{1}{(\ln\frac{\sqrt{B^2+T^2}}{T_K})^2}\frac{\pi^2}{8} (1+\frac{2
B}{T \sinh\frac{B}{T}}) \label{G2RG}
\end{equation} with the Kondo
temperature $T_K\simeq U\exp(-\pi/g_2)$. The Kondo contribution
enhances the contribution of the second channel, and gives rise to the merging of 
the "0.7" feature with the first $2e^2/h$ conductance step. As pointed out in
\cite{meir}, the resulting $T_K$ increases exponentially with
$\e_F$, in agreement with the experimental observation that $T_K$
increases exponentially with the gate voltage \cite{cron}. A similar expression
can be obtained for the total noise, but in the following we will use
the noise epression similar to that appearing in Eq.(\ref{Ginfty}), with $\hat{g_i^2}$ given by Eq.(\ref{gi}).  We also note that if
one replaces $\sqrt{B^2+T^2}$ by $\sqrt{B^2+T^2+V^2}$ in Eq.(\ref{G2RG}), the
formula (\ref{Gtot}) for the conductance not only agrees with the expansion around
large magnetic fields, but also with the expansion (in $J_i$) at small
fields \cite{appelbaum}. Thus this interpolation formula should be reliable at the whole range of magnetic
fields.

{\sl Comparison with experiment and conclusions}. 
Fig.~1 compares our calculation to the experimental results of Ref.\cite{roche} and of Ref.\cite{marcus}.
In (a) and (b) we compare the Fano factor, which is obtained, following   Ref.\cite{roche}, by subtracting from the full noise the thermal contribution (the last term in ({\ref{Ginfty}}) plus $2T[G-(e^2/h)(T_1+\hat{T}_2)]$),
and dividing this difference by the current. Plotting the Fano factor against conductance, makes the theoretical
plot practically independent of the values of $\e_0$, $U$ and $\delta$, which determine the dependence of the
conductance on gate voltage. The ratio of $g_2^2/g_1^2$ was assumed small ($=0.01$) in the spirit of the model, and
the curves for 3 values of magnetic field, in the ratio $0:3:8$ as those used in the experiment, are depicted with good agreement with experiment. The data of Ref.{\cite{marcus}} allow an even more quantitative comparison with experiment, as we used the actual values of magnetic field, voltage and temperature reported to the experiment. To get the best fit with experiment we used a $g$-factor of 0.35, indicating either the inaccuracy of the theory or the estimate of temperature in the experiment. Interestingly, the zero-field dip in the noise is quite small, even though the bare contribution of the second channel to the conductamce is negligible. This is due to the
 contributions of higher order processes, discussed below Eq.\ref{Gext}).

\vskip 0.65truecm
\begin{figure}[ht]
\begin{center}
\includegraphics [width=0.45\textwidth ]{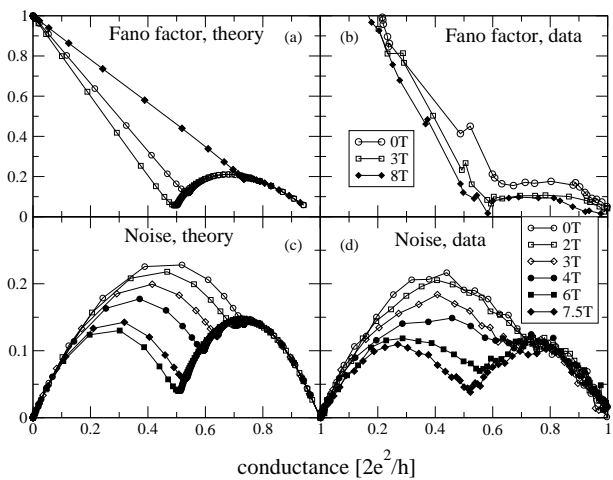}
\caption {\label{fig1} (a) The Fano factor, calculated from the theory, versus zero-bias
conductance at different magnetic fields, $g\mu_BB/k_BT=0,4.5,12$, compared to the experimental
results of Ref.\cite{roche} (b), for B=0, 3 and 8 Tesla. The parameters used in the theory were $eV=k_BT, {V^{(1)}}^2/2\pi=1, {V^{(2)}}^2/2\pi=0.01$. In (c) the noise is calculated for the same parameters as those corresponding to the data of Ref.\cite{marcus}, depicted at (d), with the magnetic field values denoted in the legend, $k_bT=280mK$ and $V=240\mu V$. The values of  ${V^{(i)}}^2$ are the same as in (a). In order to get the best comparison to the experiment a value of $g$-factor of 0.35 was used.}
\end{center}
\end{figure}

While the experiments were carried out outside the Kondo regime, due to the relative high voltage applied,
 the theory predicts that, for temperatures and voltages smaller than
the Kondo temperature, the dip in the noise
will disappear at zero field, due to the unitary limit of the Kondo
effect.

It is interesting to note that a perhaps related dip appears in the
measurement of dephasing in a quantum dot 
\cite{heiblum}, as measured by a nearby quantum point contact,
when the point contact is in the "0.7" regime. The present theory
suggests a simple explanation of this effect: as the dephasing in
the quantum dot is  by the current noise in the point contact
\cite{QDdephasing}, a dip in the noise will be associated with a
dip in the dephasing rate in the quantum dot. A detailed
calculation of this effect will be presented elsewhere \cite{tom}.

\begin{acknowledgments}
This research has been funded by the Israel Science Foundation, and by the US-Israel Binational Science Foundation. We thank the authors of Ref. \cite{marcus} for providing us the with data files.
\end{acknowledgments}

\end{document}